\documentclass[11pt]{article}

\usepackage[a4paper,margin=1in]{geometry}
\usepackage[T1]{fontenc}
\usepackage[utf8]{inputenc}
\usepackage{lmodern}
\usepackage{microtype}
\usepackage{amsmath,amssymb,amsthm,mathtools,bm}
\usepackage{graphicx}
\usepackage{physics}
\usepackage{bbm}
\usepackage{booktabs}
\usepackage{hyperref}
\usepackage[nameinlink,capitalise,noabbrev]{cleveref}

\hypersetup{
  colorlinks=true,
  linkcolor=blue,
  citecolor=blue,
  urlcolor=blue
}

\title{Microscopic Nonaffine Deformation Theory of LAOS in Polymers}

\author{
Dario Nichetti$^{1}$, Alessio Zaccone$^{2}$
}

\date{}

\begin{document}

\maketitle

\begin{center}
{\small
$^{1}$Rheonic Lab, Via Quadelle 2C, 26012, Castelleone (CR), Italy
\\
$^{2}$Department of Physics ``Aldo Pontremoli'', University of Milan, Via Celoria 16, 20133 Milan, Italy
}
\end{center}

\begin{abstract}
We develop a molecularly motivated framework connecting large-amplitude oscillatory shear (LAOS) nonlinearities in entangled polymers to frequency-dependent nonaffine relaxation in disordered solids. The central idea is that the first harmonic in LAOS measures the residual phase-locked elastic response, whereas the higher harmonics encode the Fourier signature of strain-dependent nonaffine relaxation. The finite-amplitude modulus is interpreted as a local tangent stiffness of the evolving microstructure, in the spirit of elastoplastic and incremental nonaffine models. For entangled polymers, the analogue of the decreasing coordination number in cage-breaking theories of glass mechanics is identified not with the tube-orientation tensor itself, but with the fraction of surviving tube constraints. This distinction leads naturally to a crossover description controlled by a characteristic strain amplitude \(\gamma_c\), rather than by universal fixed power-law exponents. The fitted value \(N_{\max}\simeq1.72\) indicates that the present experimental data approach a strong but not fully saturated nonlinear state, remaining below the ideal limiting value predicted for complete constraint collapse. Finally, a constraint-counting argument combining an eight-chain affine network representation with the central-force nonaffine isostatic threshold gives a limiting estimate \(|\mathrm{NLI}|_{\max}=3\). The results support the interpretation of the NLI as a Fourier-resolved dynamic nonaffinity parameter and establish a bridge between tube-based polymer dynamics, LAOS harmonic analysis, elastoplastic rheology, and microscopic nonaffine lattice dynamics.
\end{abstract}

\section{Introduction}

The rheology of entangled polymer melts and concentrated polymer solutions has traditionally been understood within the conceptual framework established by de Gennes reptation theory~\cite{deGennes1971} and the Doi--Edwards tube model~\cite{DoiEdwards1986}. In this picture, the motion of a polymer chain is constrained by surrounding chains, which collectively generate an effective confining tube. Stress relaxation then proceeds through reptation, contour-length fluctuations, chain stretch relaxation, and convective constraint release (CCR), leading to the well-known hierarchy of relaxation times and linear viscoelastic spectra characteristic of entangled polymeric systems~\cite{RubinsteinColby2003,LikhtmanMcLeish2002}.

Although enormously successful, the classical tube framework remains fundamentally mean-field in nature. The tube itself is introduced as an emergent coarse-grained topological constraint, while the microscopic structural origin of nonlinear elastic breakdown and higher-harmonic generation under large-amplitude oscillatory shear (LAOS) remains comparatively less explicit. In particular, conventional tube theory does not naturally formulate nonlinear viscoelasticity in terms of a molecular-scale loss of mechanically active load-bearing structures analogous to the cage-breaking mechanisms observed in glasses and other disordered solids.

In parallel, substantial progress has been achieved in the statistical mechanics of amorphous elasticity through the development of nonaffine lattice dynamics and microscopic theories of disordered solids~\cite{LemaitreMaloney2006, ZacconeScossaRomano2011,MilkusZaccone2017,ZacconeBook2023}. In these approaches, the elastic response is decomposed into an affine Born contribution and a nonaffine relaxation correction arising from the inability of local disordered environments to sustain purely affine deformation and, at the same time, keep mechanical equilibrium. The resulting theory naturally connects elasticity, vibrational dynamics, structural disorder, and relaxation processes at the molecular scale in amorphous materials.

A particularly important development was the nonlinear (elastoplastic) nonaffine theory of cage breaking in amorphous materials~\cite{ZacconeSchallTerentjev2014}, where the decrease of the elastic modulus under shear is traced to the progressive loss of mechanically active nearest neighbors. In this framework, nonlinear rheological response emerges directly from the strain-dependent evolution of load-bearing connectivity. Closely related ideas were later extended to colloidal glasses, where the number of long-lived neighbors and the growth of nonaffine displacements were shown to control the rheological response~\cite{LauratiEtAl2017}.

The present work is motivated by the possibility of establishing a bridge between these two conceptual frameworks: the tube-based description of polymer dynamics and the molecular-scale nonaffine description of disordered elasticity. The central idea proposed here is that the analogue of cage connectivity in an entangled polymer is the population of surviving tube constraints. Within this interpretation, nonlinear viscoelasticity arises not simply from tube orientation alone, but from the progressive loss of coherent load-bearing tube constraints during oscillatory deformation.

This perspective allows the nonlinear elastic response measured in LAOS to be interpreted in terms of a dynamic competition between:
\begin{enumerate}
    \item[(i)]
    affine-like storage associated with coherent tube orientation and surviving entanglement constraints;

    \item[(ii)]
    nonaffine relaxation associated with CCR, tube destruction, tube renewal, and nonlinear modulation of the orientational memory field.
\end{enumerate}

A particularly useful quantity in this context is the recently introduced Nonlinearity Index (NLI)~\cite{NichettiScacchi2025,ScacchiNichetti2025AJOP,NichettiScacchi2025KGK}, defined from the ratio between higher-harmonic elastic contributions and the first-harmonic storage modulus in LAOS. While originally introduced as an empirical rheological fingerprint of nonlinear viscoelasticity, the NLI admits a deeper microscopic interpretation within the present framework. Specifically, we propose that the first harmonic measures the residual coherent phase-locked elastic response, whereas the higher harmonics quantify the Fourier-resolved signature of nonlinear nonaffine relaxation.

More generally, the last two decades have seen major advances in the development of nonlinear oscillatory rheology and Fourier-transform rheology as sensitive probes of molecular and microstructural dynamics in complex fluids and soft solids~\cite{Wilhelm2002,HyunWilhelmKleinEtAl2011,EwoldtMcKinley2010,Rogers2011}. In particular, higher harmonic analysis, Lissajous--Bowditch representations, and nonlinear elastic measures have provided detailed information about strain stiffening, yielding, cage breaking, network rupture, and nonlinear relaxation processes that remain inaccessible within conventional linear viscoelasticity. These developments motivate the search for a microscopic interpretation of higher-harmonic generation in terms of evolving load-bearing microstructure and dynamic nonaffinity.

The purpose of this paper is therefore twofold. First, we develop a phenomenological mapping between the LAOS/NLI framework and the frequency-dependent nonaffine theory of disordered elasticity. Second, we extend this mapping to entangled polymers by identifying the survival fraction of tube constraints as the polymer analogue of the decreasing load-bearing coordination number in nonaffine amorphous elasticity.

This approach leads naturally to a crossover formulation for nonlinear rheology in terms of a strain-dependent competition between reptative memory retention and nonlinear tube relaxation. Rather than predicting universal fixed critical exponents, the framework predicts crossover scaling governed by the effective destruction of load-bearing tube connectivity. In this sense, the NLI emerges as a dynamic nonaffinity parameter that quantitatively measures the progressive transfer of elastic response from coherent affine-like tube elasticity into nonlinear relaxation channels.

The present framework is intentionally multiscale: nonaffine lattice dynamics provides the mechanical interpretation of relaxation channels, while tube theory supplies the molecular kinetics governing the evolution of those channels in entangled polymers.

More broadly, the theory proposed here suggests that nonlinear harmonic generation may represent a universal spectral signature of dynamically evolving load-bearing connectivity across a wide class of disordered soft materials, including entangled polymers, colloidal glasses, metallic glasses, and amorphous solids.

\section{LAOS decomposition and definition of the NLI}

We consider oscillatory shear deformation
\begin{equation}
    \gamma(t)=\gamma_0\sin(\Omega t),
\end{equation}
where \(\gamma_0\) is the strain amplitude and \(\Omega\) is the angular frequency.

In the nonlinear regime, the stress is periodic but no longer sinusoidal:
\begin{equation}
    \sigma(t)
    =
    \sum_{k=1}^{\infty}
    \left[
    \sigma'_k(\gamma_0,\Omega)\sin(k\Omega t)
    +
    \sigma''_k(\gamma_0,\Omega)\cos(k\Omega t)
    \right].
\end{equation}

For an isotropic material under symmetric shear, the elastic part contains only odd sine harmonics:
\begin{equation}
    \sigma_E(t)
    =
    \sum_{k=1,3,5,\ldots}^{\infty}
    \sigma'_k(\gamma_0,\Omega)\sin(k\Omega t).
\end{equation}

The first-harmonic storage modulus is
\begin{equation}
    G'_1(\gamma_0,\Omega)
    =
    \frac{\sigma'_1(\gamma_0,\Omega)}{\gamma_0}.
\end{equation}

The nonlinear elastic modulus used in the NLI framework is
\begin{equation}
    G'_{NL}(\gamma_0,\Omega)
    =
    \sum_{n=1}^{\infty}
    (2n+1)
    \frac{\sigma'_{2n+1}(\gamma_0,\Omega)}{\gamma_0}.
\end{equation}

Thus the generalized zero-strain elastic modulus is
\begin{equation}
    G'_0(\gamma_0,\Omega)
    =
    G'_1(\gamma_0,\Omega)
    +
    G'_{NL}(\gamma_0,\Omega),
\end{equation}
and the Nonlinearity Index is
\begin{equation}
    \mathrm{NLI}(\gamma_0,\Omega)
    =
    \frac{G'_{NL}(\gamma_0,\Omega)}
    {G'_1(\gamma_0,\Omega)}.
\end{equation}
This definition follows the generalized nonlinear elastic modulus construction introduced for nonlinear viscoelastic fingerprints in elastomers~\cite{NichettiScacchi2025}.

\section{Frequency-dependent nonaffine modulus}

In the frequency-dependent nonaffine theory of disordered solids~\cite{Palyulin,MilkusZaccone2017}, the complex modulus is written as
\begin{equation}
    G^*(\Omega)
    =
    G_{\infty}
    -
    G^*_{NA}(\Omega),
\end{equation}
where \(G_{\infty}\) is the affine high-frequency modulus and \(G^*_{NA}\) is the frequency-dependent nonaffine relaxation correction.

The storage and loss parts are
\begin{equation}
    G'(\Omega)
    =
    G_{\infty}
    -
    G'_{NA}(\Omega),
\end{equation}
and
\begin{equation}
    G''(\Omega)
    =
    G''_{NA}(\Omega).
\end{equation}

In the microscopic formulation of nonaffine viscoelasticity developed from the nonaffine lattice dynamics formalism~\cite{ZacconeScossaRomano2011,MilkusZaccone2017,ZacconeBook2023},
\begin{equation}
    G'(\Omega)
    =
    G_{\infty}
    -
    3\rho
    \int_0^{\omega_D}
    \frac{
    D(\omega)\Gamma(\omega)(\omega^2-\Omega^2)
    }
    {
    (\omega^2-\Omega^2)^2+\nu^2\Omega^2
    }
    \,d\omega,
\end{equation}
and
\begin{equation}
    G''(\Omega)
    =
    3\rho
    \int_0^{\omega_D}
    \frac{
    D(\omega)\Gamma(\omega)\nu\Omega
    }
    {
    (\omega^2-\Omega^2)^2+\nu^2\Omega^2
    }
    \,d\omega.
\end{equation}

Here \(D(\omega)\) is the vibrational density of states, \(\Gamma(\omega)\) is the nonaffine-force correlator, \(\rho\) is the number density, \(\nu\) is a damping coefficient, and \(\omega_D\) is the upper vibrational cutoff.

Thus
\begin{equation}
    G'_{NA}(\Omega)
    =
    3\rho
    \int_0^{\omega_D}
    \frac{
    D(\omega)\Gamma(\omega)(\omega^2-\Omega^2)
    }
    {
    (\omega^2-\Omega^2)^2+\nu^2\Omega^2
    }
    \,d\omega.
\end{equation}

This shows that the measured storage modulus is not the affine modulus itself, but the affine modulus after subtraction of a frequency-dependent nonaffine relaxation.

\section{LAOS extension of the nonaffine decomposition}
In LAOS, the nonaffine correction should depend on both frequency and strain amplitude:
\begin{equation}
    G'_{NA}
    =
    G'_{NA}(\gamma_0,\Omega).
\end{equation}

The finite-amplitude extension of the shear modulus should be interpreted in the spirit of elastoplastic and incremental  models~\cite{Nicolas2018}, where the shear modulus is understood as a local or tangent modulus defined along the deformation trajectory. In this picture, the stress evolution obeys an incremental constitutive relation of the form
\begin{equation}
\sigma(\gamma_i)
=
\sigma(\gamma_{i-1})
+
G(\gamma_i)
\left(
\gamma_i-\gamma_{i-1}
\right),
\end{equation}
as discussed for example in nonaffine elastoplastic formulations and mesoscopic yielding models~\cite{Ivan,Tanguy}. The strain-dependent modulus \(G(\gamma)\) therefore characterizes the instantaneous load-bearing stiffness of the evolving microstructure rather than a single equilibrium linear-response modulus.

The first-harmonic LAOS modulus is identified with the first-harmonic projection of this effective modulus:
\begin{equation}
    G'_1(\gamma_0,\Omega)
    \simeq
    G_{\infty}(\gamma_0)
    -
    G'_{NA,1}(\gamma_0,\Omega).
\end{equation}

Thus \(G'_1\) is not purely affine. It is the residual coherent, phase-locked elastic response after nonaffine relaxation.
Similarly, the nonlinear modulus \(G'_{NL}\) should not be identified directly with the nonaffine modulus. Rather, it is the higher-harmonic projection of the strain-dependent nonaffine relaxation:
\begin{equation}
    G'_{NL}(\gamma_0,\Omega)
    \simeq
    -
    G'_{NA,h}(\gamma_0,\Omega),
\end{equation}
where \(G'_{NA,h}\) denotes the higher-harmonic part of the nonaffine correction.

Therefore,
\begin{equation}
    G'_0(\gamma_0,\Omega)
    =
    G'_1(\gamma_0,\Omega)
    +
    G'_{NL}(\gamma_0,\Omega)
\end{equation}
becomes
\begin{equation}
    G'_0(\gamma_0,\Omega)
    \simeq
    G_{\infty}(\gamma_0)
    -
    \left[
    G'_{NA,1}(\gamma_0,\Omega)
    +
    G'_{NA,h}(\gamma_0,\Omega)
    \right].
\end{equation}

Hence
\begin{equation}
    G'_0(\gamma_0,\Omega)
    \simeq
    G_{\infty}(\gamma_0)
    -
    G'_{NA,\mathrm{tot}}(\gamma_0,\Omega).
\end{equation}

This is the LAOS analogue of the nonaffine decomposition. In polymer systems, the nonaffine relaxation channel should not be interpreted literally as instantaneous local force-balance restoration, as in atomic glasses, but rather as the dynamically heterogeneous loss of coherent tube-memory response under deformation.

\section{Interpretation of the NLI}

Substitution into the NLI gives
\begin{equation}
    \mathrm{NLI}(\gamma_0,\Omega)
    =
    \frac{G'_{NL}(\gamma_0,\Omega)}
    {G'_1(\gamma_0,\Omega)}
    \simeq
    -
    \frac{
    G'_{NA,h}(\gamma_0,\Omega)
    }
    {
    G_{\infty}(\gamma_0)-G'_{NA,1}(\gamma_0,\Omega)
    }.
\end{equation}
The NLI framework and generalized nonlinear elastic modulus were recently introduced as sensitive rheological fingerprints for nonlinear viscoelasticity and polymer--filler interactions in filled elastomers~\cite{NichettiScacchi2025}.

Thus the NLI is not simply the ratio of a nonaffine modulus to an affine modulus. It is the ratio between the higher-harmonic nonlinear part of the nonaffine relaxation channel and the residual first-harmonic storage modulus:
\begin{equation}
    \mathrm{NLI}
    \simeq
    -
    \frac{
    \text{higher-harmonic nonaffine relaxation}
    }
    {
    \text{residual first-harmonic elastic storage}
    }.
\end{equation}

\section{Connectivity loss in glasses and tube survival in polymers}

In the nonaffine theory of amorphous solids, the shear modulus of a central-force network scales as
\begin{equation}
    G(\gamma)
    =
    C\left[n_b(\gamma)-n_b^c\right],
\end{equation}
where \(n_b(\gamma)\) is the number of mechanically active neighbors and \(n_b^c\) is the critical connectivity for rigidity.

In the nonlinear cage-breaking theory of amorphous solids~\cite{ZacconeSchallTerentjev2014,Zaccone_2020}, shear progressively removes mechanically-active neighbors:
\begin{equation}
    n_b(\gamma)
    =
    \frac{n_b^0}{2}
    \left(
    1+e^{-A\gamma}
    \right),
\end{equation}
with
\begin{equation}
    A
    =
    \frac{T_g}{T}
    +
    \frac{1}{\dot{\gamma}\tau_c}.
\end{equation}

Thus increasing strain decreases the load-bearing connectivity. The modulus decreases when the remaining connectivity can no longer sustain affine elastic deformation.

For colloidal glasses, the analogous structural variable is the number of long-lived neighbors~\cite{LauratiEtAl2017}:
\begin{equation}
    n(\gamma)
    =
    (n_0-n_{\infty})
    e^{-(\gamma/\xi)^2}
    +
    n_{\infty}.
\end{equation}

The decrease of \(n(\gamma)\) is accompanied by increasing local nonaffine displacement:
\begin{equation}
    u^2_{NA}(\gamma)
    \simeq
    a+b^2\gamma^2+2bc\gamma^3.
\end{equation}

Therefore, in glasses and colloidal glasses, the central structural mechanism is the loss of long-lived load-bearing neighbors.

\section{Polymer analogue: tube survival and tube orientation}

For an entangled polymer melt, the relevant microstructural object is not an atomic cage but the entanglement tube. The elastic stress is controlled by the orientation tensor of primitive-path segments,
\begin{equation}
    \bm{S}
    =
    \langle \bm{u}\bm{u}\rangle,
\end{equation}
where \(\bm{u}\) is the unit tangent vector along the primitive path.

The shear component of the tube stress may be written schematically as
\begin{equation}
    \sigma^E_{xy}
    =
    G_e^0\,
    q(\gamma,t)\,
    \lambda^2(\gamma,t)\,
    S_{xy}(\gamma,t),
\end{equation}
where
\begin{equation}
    S_{xy}=\langle u_xu_y\rangle.
\end{equation}

Here \(q(\gamma,t)\) is the fraction of surviving tube constraints, \(\lambda\) is the chain stretch ratio, and \(G_e^0\) is the equilibrium tube modulus.

The key correspondence is
\begin{equation}
    \frac{n_b(\gamma)-n_b^c}{n_b^0-n_b^c}
    \quad
    \longleftrightarrow
    \quad
    q(\gamma,t).
\end{equation}

Thus \(q\), not \(S_{xy}\), is the direct analogue of the decreasing load-bearing connectivity \(n_b\). The tensor \(S_{xy}\) describes the orientational state of the surviving constraints.

The stress-level analogy is
\begin{equation}
    \sigma_E^{glass}
    \sim
    K\left[n_b(\gamma)-n_b^c\right]\gamma,
\end{equation}
and
\begin{equation}
    \sigma_E^{tube}
    =
    G_e^0
    q(\gamma,t)
    \lambda^2(\gamma,t)
    S_{xy}(\gamma,t).
\end{equation}

\section{Linear tube spectrum as the reference state}

Before discussing nonlinear LAOS, it is useful to identify the linear viscoelastic reference state. In the Doi--Edwards tube picture~\cite{DoiEdwards1986,deGennes1971}, stress relaxation is governed by the survival of tube-orientation memory. In the simplest single-time approximation, the shear component of the orientation tensor obeys
\begin{equation}
    \frac{dS_{xy}}{dt}
    =
    \frac{1}{3}\dot{\gamma}
    -
    \frac{S_{xy}}{\tau_d},
\end{equation}
where \(\tau_d\) is the disengagement or terminal reptation time.

For sinusoidal shear,
\begin{equation}
    \gamma(t)=\gamma_0\sin(\omega t),
    \qquad
    \dot{\gamma}(t)=\gamma_0\omega\cos(\omega t),
\end{equation}
the steady-periodic solution is
\begin{equation}
    S_{xy}(t)
    =
    \frac{\gamma_0}{3}
    \frac{\omega^2\tau_d^2}{1+\omega^2\tau_d^2}
    \sin(\omega t)
    +
    \frac{\gamma_0}{3}
    \frac{\omega\tau_d}{1+\omega^2\tau_d^2}
    \cos(\omega t).
\end{equation}

Since
\begin{equation}
    \sigma_{xy}(t)=G_N^0 S_{xy}(t),
\end{equation}
one obtains
\begin{equation}
    G'(\omega)
    =
    \frac{G_N^0}{3}
    \frac{\omega^2\tau_d^2}{1+\omega^2\tau_d^2},
\end{equation}
and
\begin{equation}
    G''(\omega)
    =
    \frac{G_N^0}{3}
    \frac{\omega\tau_d}{1+\omega^2\tau_d^2}.
\end{equation}

Thus the elastic and viscous moduli are not imposed phenomenologically. They emerge from the same molecular memory equation: the affine term builds recoverable tube orientation, whereas the relaxation term erases tube memory. In this sense,
\begin{equation}
    \text{elasticity}
    =
    \text{molecular memory retained},
\end{equation}
whereas
\begin{equation}
    \text{viscosity}
    =
    \text{molecular memory lost per unit time}.
\end{equation}

The usual linear tube spectrum contains several characteristic time scales associated with local segmental relaxation, entanglement dynamics, Rouse relaxation, and terminal reptation~\cite{DoiEdwards1986,RubinsteinColby2003,LikhtmanMcLeish2002,CaoLikhtman2015}:
\begin{equation}
    \tau_0 < \tau_e < \tau_R < \tau_d,
\end{equation}
corresponding respectively to local segmental motion, entanglement-scale relaxation, Rouse relaxation of the whole chain, and terminal reptation. The nonlinear LAOS problem should therefore be interpreted as a strain-amplitude-dependent modification of this pre-existing molecular spectrum.

\section{Convective constraint release as tube-cage breaking}

Convective constraint release (CCR), originally introduced by Marrucci as a mechanism for flow-induced release of entanglement constraints in polymer melts \cite{Marrucci1996,MarrucciIanniruberto1996}, is the polymer analogue of cage breaking and plays a central role in modern tube-theory descriptions of nonlinear entangled polymer rheology, including the Rolie--Poly framework \cite{LikhtmanGraham2003}. In glasses, shear removes neighbors from the cage. In glasses, shear removes neighbors from the cage. In entangled polymers, Marrucci's CCR mechanism describes how flow convects surrounding chains and progressively releases the topological constraints defining the confining tube. In entangled polymers, flow convects surrounding chains and releases the constraints defining the tube. This mechanism plays a central role in modern tube models of nonlinear polymer rheology~\cite{DoiEdwards1986,LikhtmanMcLeish2002,CaoLikhtman2015}.

A minimal evolution equation for the tube survival fraction is
\begin{equation}
    \frac{dq}{dt}
    =
    -
    \frac{q}{\tau_d}
    -
    \beta |\dot{\gamma}|q
    +
    R_q,
\end{equation}
where \(\tau_d\) is the reptation time, \(\beta\) is the CCR efficiency, and \(R_q\) represents constraint reformation.

In the strong-flow limit,
\begin{equation}
    \frac{dq}{dt}
    \simeq
    -
    \beta |\dot{\gamma}|q.
\end{equation}

For monotonic shear,
\begin{equation}
    q(\gamma)
    \simeq
    e^{-\beta\gamma}.
\end{equation}

This has the same structure as cage connectivity loss:
\begin{equation}
    n_b(\gamma)-n_b^{\infty}
    \propto
    e^{-A\gamma}.
\end{equation}

Thus, at the level of structural relaxation rates,
\begin{equation}
    A
    \quad
    \longleftrightarrow
    \quad
    \beta.
\end{equation}

Under oscillatory shear,
\begin{equation}
    \dot{\gamma}(t)
    =
    \gamma_0\Omega\cos(\Omega t),
\end{equation}
so the CCR rate is
\begin{equation}
    \tau^{-1}_{CCR}(t)
    =
    \beta |\dot{\gamma}(t)|
    =
    \beta\gamma_0\Omega |\cos(\Omega t)|.
\end{equation}

Tube constraints are therefore depleted most strongly near zero-strain crossings, where the shear rate is maximal. This intra-cycle modulation is the microscopic origin of higher harmonic generation in this framework.

\section{Crossover strain for onset of nonlinear nonaffinity}
The crossover strain \(\gamma_c\) should be interpreted relative to the linear tube spectrum. In linear rheology, the material crosses from terminal viscous response to rubbery plateau response as \(\omega\tau_d\) increases, and from plateau response toward Rouse/glassy regimes as \(\omega\tau_R\), \(\omega\tau_e\), and \(\omega\tau_0\) become relevant. In LAOS, increasing \(\gamma_0\) at fixed \(\omega\) activates nonlinear relaxation channels, especially CCR and tube dilation, so that the effective relaxation time becomes strain dependent.

The experimental trends suggest that the NLI is not governed by a universal power law over the full strain range. Instead, there is a crossover near a characteristic strain amplitude \(\gamma_c\).

\begin{equation}
    \tau_{\mathrm{eff}}^{-1}
    =
    \tau_d^{-1}
    +
    \tau_{CCR}^{-1},
    \qquad
    \tau_{CCR}^{-1}
    \sim
    \beta |\bm{\kappa}:\bm{S}|.
\end{equation}

For simple shear,
\begin{equation}
    \bm{\kappa}:\bm{S}
    =
    \dot{\gamma}S_{xy}.
\end{equation}

Hence, under LAOS,
\begin{equation}
    \tau_{CCR}^{-1}
    \sim
    \beta \gamma_0\Omega |S_{xy}(t)\cos(\Omega t)|.
\end{equation}

Using \(S_{xy}=O(1)\) in the strongly oriented nonlinear regime gives the estimate
\begin{equation}
    \gamma_c
    \sim
    \frac{1}{\beta\Omega\tau_d}.
\end{equation}

If instead the onset occurs while \(S_{xy}\sim\gamma_0\), then the crossover condition becomes
\begin{equation}
    \beta \Omega \tau_d \gamma_c^2 \sim 1,
\end{equation}
so that
\begin{equation}
    \gamma_c
    \sim
    (\beta\Omega\tau_d)^{-1/2}.
\end{equation}
Here \(\gamma_c\) should be interpreted as the crossover value in the normalized strain variable
$\frac{\gamma_0}{\gamma_0^{LR}}$,
rather than as an absolute strain amplitude.

Thus the experimentally observed onset strain can distinguish whether CCR begins in a weak-orientation regime or in an already strongly oriented tube regime.

For \(\gamma_0\ll\gamma_c\), reptation dominates, tube constraints remain coherent over the cycle, and the response is approximately affine and first-harmonic dominated. This physical picture is also consistent with startup shear experiments and molecular simulations of entangled melts, where stress overshoots emerge once the imposed deformation rate exceeds the characteristic reptation rate~\cite{CaoLikhtman2015}.

For \(\gamma_0\gtrsim\gamma_c\), CCR becomes strong enough to destroy phase-locked tube coherence within each cycle. Higher harmonics then grow rapidly.

For \(\gamma_0\gg\gamma_c\), the response enters a strongly nonlinear constraint-collapse regime.

Therefore the natural scaling form is
\begin{equation}
    \mathrm{NLI}(\gamma_0,\Omega)
    =
    \mathcal{F}
    \left(
    \frac{\gamma_0}{\gamma_c}
    \right).
\end{equation}

The limiting behavior is
\begin{equation}
    \mathcal{F}(x)\simeq 0,
    \qquad
    x\ll1,
\end{equation}
and
\begin{equation}
    \mathcal{F}(x)\sim x^m,
    \qquad
    x\gg1.
\end{equation}

The exponent \(m\) should be regarded as an effective crossover exponent, not as a universal critical exponent.

\section{Effective power-law exponents}

In the strong nonlinear regime, write
\begin{equation}
    q_0(\gamma_0)\sim\gamma_0^{-\alpha_q},
\end{equation}
\begin{equation}
    \lambda_0^2(\gamma_0)\sim\gamma_0^{\alpha_\lambda},
\end{equation}
and
\begin{equation}
    S_{xy,1}(\gamma_0)\sim\gamma_0^{\alpha_u},
\end{equation}
where \(S_{xy,1}\) is the first-harmonic projection of the orientation tensor.

The first elastic harmonic scales as
\begin{equation}
    \sigma'_1
    \sim
    G_e^0
    \gamma_0^{-\alpha_q}
    \gamma_0^{\alpha_\lambda}
    \gamma_0^{\alpha_u}.
\end{equation}

Therefore
\begin{equation}
    G'_1
    =
    \frac{\sigma'_1}{\gamma_0}
    \sim
    \gamma_0^{-1-\alpha_q+\alpha_\lambda+\alpha_u}.
\end{equation}

Let the weighted higher-harmonic stress scale as
\begin{equation}
    \sigma'_h
    \equiv
    \sum_{n=1}^{\infty}
    (2n+1)\sigma'_{2n+1}
    \sim
    \gamma_0^{\alpha_h}.
\end{equation}

Then
\begin{equation}
    G'_{NL}
    =
    \frac{\sigma'_h}{\gamma_0}
    \sim
    \gamma_0^{\alpha_h-1}.
\end{equation}

Thus
\begin{equation}
    \mathrm{NLI}
    =
    \frac{G'_{NL}}{G'_1}
    \sim
    \gamma_0^{m},
\end{equation}
with
\begin{equation}
    m
    =
    \alpha_h+\alpha_q-\alpha_\lambda-\alpha_u.
\end{equation}

Linear NLI growth,
\begin{equation}
    \mathrm{NLI}\sim\gamma_0,
\end{equation}
requires
\begin{equation}
    \alpha_h+\alpha_q-\alpha_\lambda-\alpha_u=1.
\end{equation}

This condition expresses a balance between nonlinear harmonic generation and coherent tube elasticity.

\section{Physical interpretation of the exponent condition}

The exponent \(\alpha_h\) measures the growth of nonlinear harmonic stress. It quantifies intra-cycle waveform distortion caused by nonlinear tube relaxation.

The exponent \(\alpha_q\) measures the decay of surviving tube constraints:
\begin{equation}
    q(\gamma_0)\sim\gamma_0^{-\alpha_q}.
\end{equation}

The exponents \(\alpha_\lambda\) and \(\alpha_u\) stabilize the first harmonic. The exponent \(\alpha_\lambda\) measures stretch amplification:
\begin{equation}
    \lambda^2\sim\gamma_0^{\alpha_\lambda},
\end{equation}
while \(\alpha_u\) measures coherent phase-locked orientation:
\begin{equation}
    S_{xy,1}\sim\gamma_0^{\alpha_u}.
\end{equation}

Therefore
\begin{equation}
    \alpha_h+\alpha_q-\alpha_\lambda-\alpha_u=1
\end{equation}
means that nonlinear relaxation and tube-constraint loss exceed coherent stretch-orientation storage by exactly one power of strain amplitude.

\section{Regime classification}

The experimentally measured slope
\begin{equation}
    m
    =
    \frac{d\log(\mathrm{NLI})}{d\log\gamma_0}
\end{equation}
can be used to classify nonlinear tube dynamics:
\begin{equation}
    m<1
    \quad\Rightarrow\quad
    \text{persistent affine-like elasticity or stretch hardening},
\end{equation}
\begin{equation}
    m\simeq1
    \quad\Rightarrow\quad
    \text{CCR-dominated constraint-collapse regime},
\end{equation}
\begin{equation}
    m>1
    \quad\Rightarrow\quad
    \text{rapid tube dilution or fragile entanglement structure}.
\end{equation}

A compact summary is given in \cref{tab:regimes}.

\begin{table}[ht]
\centering
\begin{tabular}{llll}
\toprule
Regime & \(G'_1\) & \(G'_{NL}\) & \(\mathrm{NLI}\) \\
\midrule
Weakly nonlinear tube & \(\gamma_0^0\) & small & no robust power law \\
CCR stress saturation & \(\gamma_0^{-1}\) & \(\gamma_0^0\) & \(\gamma_0^1\) \\
Tube dilution + CCR & \(\gamma_0^{-1-\alpha_q}\) & \(\gamma_0^0\) & \(\gamma_0^{1+\alpha_q}\) \\
Stretch dominated & \(\gamma_0^{-1+\alpha_\lambda}\) & \(\gamma_0^0\) & \(\gamma_0^{1-\alpha_\lambda}\) \\
\bottomrule
\end{tabular}
\caption{Effective scaling regimes for the NLI. These are crossover regimes, not universal critical exponents.}
\label{tab:regimes}
\end{table}

\section{Molecular architecture and saturation level}

The crossover function
\begin{equation}
    \mathrm{NLI}
    =
    \mathcal{F}
    \left(
    \frac{\gamma_0}{\gamma_c}
    \right)
\end{equation}
also provides a natural interpretation of the difference between linear and branched polymers.

Linear polymers can undergo stronger tube orientation loss and stronger harmonic distortion once \(\gamma_0\gtrsim\gamma_c\). Therefore their NLI may approach values of order unity.

Branching modifies the survival and relaxation kinetics of entanglement constraints, thereby altering the nonlinear nonaffine relaxation channel detected through the higher harmonics.

Branched polymers preserve tube coherence more effectively because branch points act as additional topological constraints. This can reduce CCR efficiency, increase the effective crossover strain, and suppress the asymptotic NLI level:
\begin{equation}
    \gamma_c^{branched}
    >
    \gamma_c^{linear},
\end{equation}
and
\begin{equation}
    \mathcal{F}_{branched}(x)
    <
    \mathcal{F}_{linear}(x)
    \qquad
    \text{at comparable }x.
\end{equation}

Thus the NLI can be interpreted as a normalized dynamic nonaffinity parameter: low NLI indicates coherent affine-like tube response, whereas high NLI indicates strong nonlinear tube relaxation and loss of phase-locked orientation.

\section{Comparison with experimental data}

To test the proposed crossover interpretation, we fitted the experimental master curve of the nonlinear index reported for polymer melts and elastomers in the NLI framework~\cite{NichettiScacchi2025}, using the phenomenological crossover function predicted by the present framework.

The experimental quantity is the normalized nonlinear elastic contribution
\begin{equation}
    \mathrm{NLI}
    =
    -\frac{G'_{NL}}{G'},
\end{equation}
plotted as a function of the normalized strain amplitude
\begin{equation}
    \frac{\gamma_0}{\gamma_0^{LR}},
\end{equation}
where \(\gamma_0^{LR}\) denotes the onset strain for departure from linear rheology.

The theoretical framework developed above suggests that the nonlinear response is governed by a crossover between a low-strain regime dominated by coherent affine-like tube elasticity and a nonlinear regime dominated by CCR-induced tube relaxation and higher-harmonic generation.

The simplest interpolation consistent with these limits is
\begin{equation}
\label{eq:NLI_crossover_fit}
    \mathrm{NLI}(\gamma_0)
    =
    N_{\max}
    \frac{
    \left(\gamma_0/\gamma_c\right)^m
    }
    {
    1+
    \left(\gamma_0/\gamma_c\right)^m
    }.
\end{equation}

Here:
\begin{enumerate}
    \item[(i)]
    \(N_{\max}\) is the asymptotic saturation level of nonlinear elastic distortion;

    \item[(ii)]
    \(\gamma_c\) is the crossover strain amplitude for onset of strong nonlinear tube relaxation;

    \item[(iii)]
    \(m\) is an effective crossover exponent describing how sharply nonlinear harmonics emerge.
\end{enumerate}

This fitting form follows directly from the crossover scaling relation
\begin{equation}
    \mathrm{NLI}
    =
    \mathcal{F}
    \left(
    \frac{\gamma_0}{\gamma_c}
    \right),
\end{equation}
derived above.

Indeed:
\begin{equation}
    \mathrm{NLI}
    \rightarrow
    0,
    \qquad
    \gamma_0\ll\gamma_c,
\end{equation}
while
\begin{equation}
    \mathrm{NLI}
    \rightarrow
    N_{\max},
    \qquad
    \gamma_0\gg\gamma_c.
\end{equation}

In the intermediate crossover regime,
\begin{equation}
    \mathrm{NLI}
    \sim
    \left(
    \frac{\gamma_0}{\gamma_c}
    \right)^m,
\end{equation}
which reproduces the effective power-law behavior discussed previously.

\begin{figure}[h]
\centering
\includegraphics[width=0.8\textwidth]{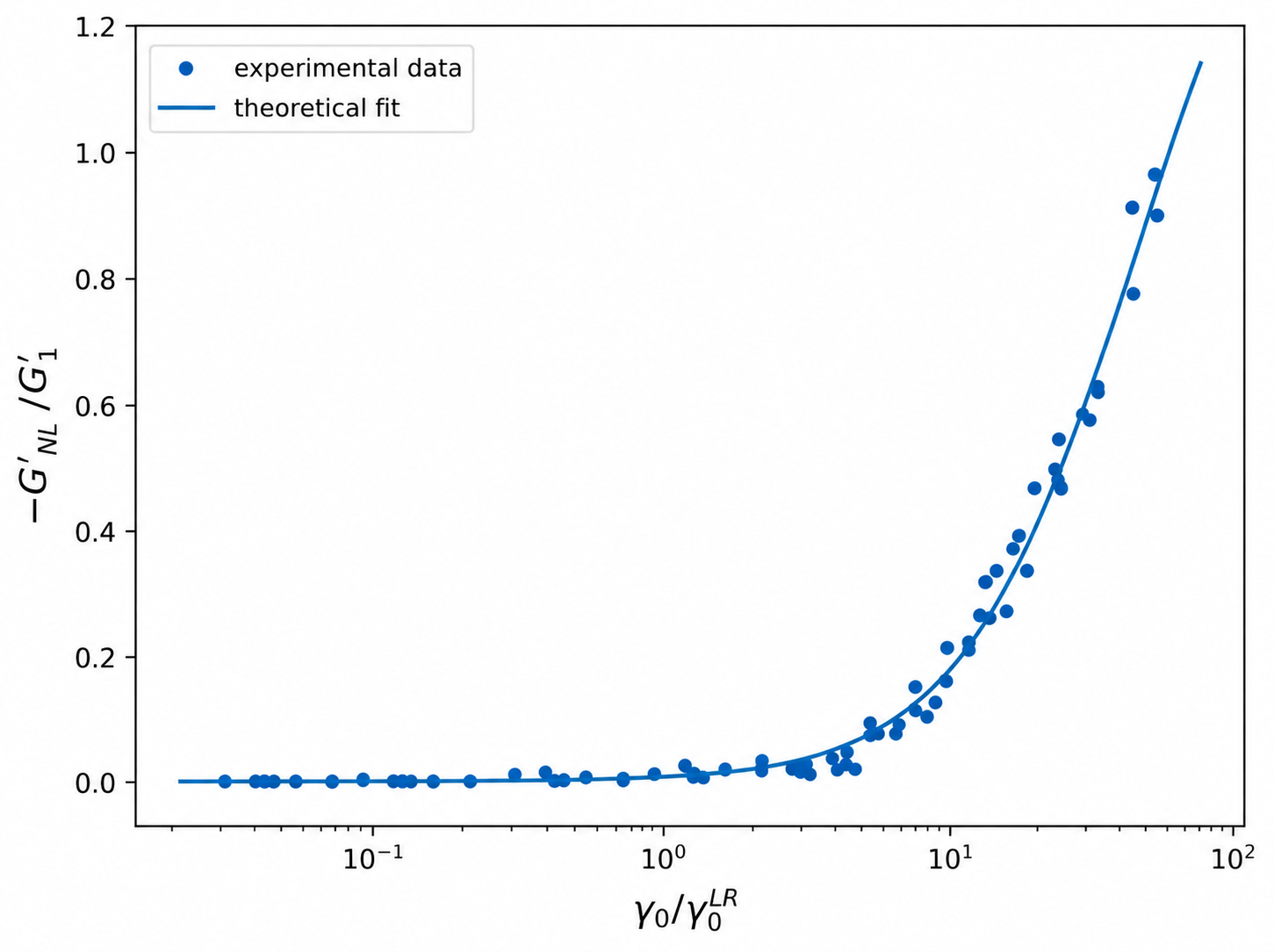}
\caption{
Experimental master curve of the normalized nonlinear elastic contribution,
\(\mathrm{NLI}=-G'_{NL}/G'_1\), plotted as a function of the normalized strain amplitude
\(\gamma_0/\gamma_0^{LR}\). Symbols are experimental data from the literature \cite{NichettiScacchi2025}, while the solid line is the theoretical crossover fit given by Eq.~(\ref{eq:NLI_crossover_fit}).
The data include polymers with markedly different molecular architectures, including predominantly linear linear-low-density polyethylene (LLDPE), highly branched low-density polyethylene (LDPE), randomly branched emulsion-polymerized styrene--butadiene rubber (e-SBR), and solution-polymerized partially coupled styrene--butadiene rubber (s-SBR). Despite these architectural differences, the normalized generalized modulus exhibits an approximately universal crossover from linear to nonlinear viscoelasticity.}
\label{fig:nli_fit}
\end{figure}

The resulting fit is shown in Fig. \cref{fig:nli_fit}.

The fitted quantity in Fig.~\ref{fig:nli_fit} is the experimental nonlinear index
\[
\mathrm{NLI}
=
-\frac{G'_{NL}}{G'_1},
\]
namely the ratio between the higher-harmonic nonlinear elastic contribution and the residual first-harmonic storage modulus.
The fitted parameters are summarized in \cref{tab:fitparams}.

It is important to stress that the fitted value \(N_{\max}\simeq 1.72\) is not imposed a priori and should not be confused with the theoretical upper bound \(|\mathrm{NLI}|_{\max}=3\) derived below as an upper bound for the unnormalized nonlinear index. Rather, it is the maximum nonlinear level reached by the present experimental data within the accessible LAOS window. The fact that \(N_{\max}\) is of order two indicates that the measured systems approach a strongly nonlinear state, but do not yet reach the ideal fully saturated constraint-collapse limit.

The crossover parameter \(\gamma_c\simeq39.2\) is expressed in the normalized strain variable \(\gamma_0/\gamma_0^{LR}\). Therefore, \(\gamma_c\) should be interpreted as the strain amplitude, measured relative to the material-specific linear-regime limit \(\gamma_0^{LR}\), at which nonlinear tube relaxation becomes dominant.

\begin{table}[ht]
\centering
\begin{tabular}{lc}
\toprule
Parameter & Fitted value \\
\midrule
\(N_{\max}\) & \(1.72\) \\
\(\gamma_c\) & \(39.2\) \\
\(m\) & \(1.58\) \\
\bottomrule
\end{tabular}
\caption{
Fitted parameters for the crossover model using the data points of various polymers collected in \cite{NichettiScacchi2025}. Here \(N_{\max}\) denotes the maximum nonlinear level reached within the experimentally accessible LAOS window, while \(\gamma_c\) is expressed in the normalized strain variable \(\gamma_0/\gamma_0^{LR}\). The fit gives \(R^2=0.988\).
}
\label{tab:fitparams}
\end{table}

The apparent collapse of the normalized modulus curves indicates that the reduction of coherent first-harmonic elasticity follows a broadly universal nonlinear crossover. However, as shown later through the NLI analysis, the higher-harmonic nonlinear response remains highly sensitive to molecular topology and branching architecture.

Several important physical conclusions emerge from the fit.

First, the fitted saturation value
\begin{equation}
    N_{\max}\approx1.72
\end{equation}
should not be interpreted as the absolute asymptotic upper bound of the nonlinear response. Rather, it represents the maximum nonlinear level reached within the experimentally accessible LAOS window. Physically, this indicates that, in the strongly nonlinear regime explored experimentally, the nonlinear elastic contribution becomes comparable in magnitude to the residual first-harmonic storage modulus. A substantial fraction of the elastic response is therefore transferred from coherent affine-like storage into nonlinear harmonic distortion generated by tube relaxation. The larger theoretical limit
\begin{equation}
    |\mathrm{NLI}|_{\max}=3
\end{equation}
derived below corresponds instead to an ideal fully saturated nonlinear state.

Second, the crossover parameter
\begin{equation}
    \gamma_c\approx 40
\end{equation}
identifies the onset of strong nonlinear tube dynamics in the normalized strain variable
\begin{equation}
    \frac{\gamma_0}{\gamma_0^{LR}},
\end{equation}
where \(\gamma_0^{LR}\) denotes the material-dependent onset strain for departure from linear viscoelasticity.

Within the present framework, this corresponds to the regime where CCR becomes comparable to or larger than reptative memory retention:
\begin{equation}
    \tau_{CCR}^{-1}
    \sim
    \tau_d^{-1}.
\end{equation}

Thus the fit supports the interpretation that the observed nonlinear harmonic growth is controlled by progressive destruction of phase-locked tube coherence.

Third, the value \(m\simeq1.58\) should be interpreted as an effective crossover exponent, not as a universal critical exponent. It indicates that the onset of nonlinear relaxation is sharper than the simplest CCR-saturation scenario, for which one would expect \(m\simeq1\). In the exponent balance
\[
m=\alpha_h+\alpha_q-\alpha_\lambda-\alpha_u,
\]
a value close to \(3/2\) suggests that higher-harmonic generation and constraint loss are reinforced by an additional nonlinear amplification mechanism, such as progressive tube dilution, loss of orientational memory, or increasingly heterogeneous constraint release during the LAOS cycle.

Physically, this suggests a genuine nonlinear constraint-collapse regime characterized by:
\begin{enumerate}
    \item[(i)]
    strong CCR activity;

    \item[(ii)]
    rapid loss of surviving tube constraints;

    \item[(iii)]
    progressive destruction of phase-locked orientation memory;

    \item[(iv)]
    accelerated higher-harmonic generation.
\end{enumerate}

The fit therefore supports the central interpretation proposed in this work: the nonlinear index behaves as a Fourier-resolved dynamic nonaffinity parameter that measures the progressive transfer of elastic response from coherent affine-like tube elasticity into nonlinear nonaffine relaxation channels.

Finally, the existence of a successful master-curve fit strongly supports the crossover formulation
\begin{equation}
    \mathrm{NLI}
    =
    \mathcal{F}
    \left(
    \frac{\gamma_0}{\gamma_c}
    \right),
\end{equation}
rather than the existence of a universal asymptotic power law valid over all strain amplitudes.

\section{Constraint-counting estimate of the limiting NLI=3}

We now derive a possible upper bound for the nonlinear index (NLI) using the nonaffine deformation framework with network constraint-counting argument.

Within the nonaffine deformation framework, the elastic modulus is written as \cite{Zaccone2011}:
\begin{equation}
    G
    =
    G_A-G_{NA},
\end{equation}
where \(G_A\) is the affine elastic contribution and \(G_{NA}\) is the nonaffine relaxation correction. In nonaffine lattice dynamics, the affine modulus scales with the number of mechanically active load-bearing constraints, while the nonaffine correction scales with the number of constraints required to absorb nonaffine relaxation. Schematically \cite{Zaccone2011,Zaccone2013,ZacconeTerentjev2013,ZacconeBook2023},
\begin{equation}
    G_A\sim n_b,
    \qquad
    G_{NA}\sim n_b^{c},
\end{equation}
so that
\begin{equation}
    G
    \sim
    (n_b-n_b^{c}).
\end{equation}

At maximum nonlinear saturation, we denote the corresponding critical nonaffine coordination by \(z_{c,\max}\), and write
\begin{equation}
    G_{\min}
    =
    G_A-G_{NA,\max}
    \sim
    (n_b-n_{b,\max}^{c}).
\end{equation}

The fraction of affine elasticity transferred into nonaffine relaxation channels is therefore
\begin{equation}
    \chi_{\max}
    =
    \frac{G_{NA,\max}}{G_A}
    =
    \frac{n_{b,\max}^{c}}{n_b}.
\end{equation}

Since the nonlinear index measures the nonlinear nonaffine contribution relative to the residual coherent first-harmonic modulus,
\begin{equation}
    |\mathrm{NLI}|_{\max}
    =
    \frac{G_{NA,\max}}
    {G_A-G_{NA,\max}},
\end{equation}
one obtains
\begin{equation}
    |\mathrm{NLI}|_{\max}
    =
    \frac{\chi_{\max}}
    {1-\chi_{\max}}
    =
    \frac{n_{b,\max}^{c}}
    {n_b-n_{b,\max}^{c}}.
\end{equation}

The key issue is then the physical interpretation of \(n_b\) and \(n_{b,\max}^{c}\).

For flexible chains (where the bond-bending rigidity of the covalent bonds is negligible), we take
\begin{equation}
    n_{b,\max}^{c}=2d=6,
\end{equation}
namely the Maxwell isostatic threshold for a three-dimensional ($d=3$) central-force elastic network, as derived in \cite{Zaccone2011}. This choice corresponds to the limiting case in which the mechanically active constraints behave effectively as central-force constraints and bond-bending rigidity is absent or fully relaxed.

This is distinct from the polymer-glass constraint-counting theory of Zaccone and Terentjev~\cite{ZacconeTerentjev2013} and subsequent work \cite{LappalaZacconeTerentjev2016}, where covalent chain connectivity and bond-bending constraints reduce the marginal rigidity threshold to $z^\ast\simeq4$
for long chains. In the present context, however, we are interested in the limiting nonlinear state of a fully flexible entangled network because this is what provides the upper bound for $G_{NA}$, because $|G_{NA}| \sim 4$ for stiff chains with bond-bending \cite{ZacconeTerentjev2013,LappalaZacconeTerentjev2016}, while $|G_{NA}| \sim 6$ for fully-flexible chains (no bond-bending) \cite{Zaccone2011}, where the relevant nonaffine threshold reverts to the central-force value \(n_{b,\max}^{c}=6\).

To estimate the mechanically-active coordination \(n_b\) of the polymer melt, we invoke the eight-chain representation used in the Arruda--Boyce and Palmer--Boyce models of rubber elasticity~\cite{ArrudaBoyce1993,PalmerBoyce2008}. In these models, the representative volume element contains eight load-bearing chains oriented along the body diagonals of a cube. This gives a natural effective mechanical network coordination
\begin{equation}
    n_b=8.
\end{equation}

Importantly, this quantity should not be interpreted directly as the microscopic number of nearest-neighbour monomer contacts. Rather, it represents the number of coarse-grained load-bearing elastic strands in the affine network cell. In this sense, the Arruda--Boyce eight-chain construction provides a natural affine coordination compatible with the nonaffine deformation framework.

Substituting
\begin{equation}
    n_b=8,
    \qquad
    n_{b,\max}^{c}=6
\end{equation}
into the constraint-counting expression yields
\begin{equation}
    |\mathrm{NLI}|_{\max}
    =
    \frac{6}{8-6}
    =
    3.
\end{equation}

Equivalently,
\begin{equation}
    \chi_{\max}
    =
    \frac{n_{b,\max}^{c}}{n_b}
    =
    \frac{6}{8}
    =
    \frac{3}{4},
\end{equation}
leaving a residual coherent affine-like fraction
\begin{equation}
    1-\chi_{\max}
    =
    \frac{1}{4}.
\end{equation}

Thus, in the maximally nonlinear state, three quarters of the affine elastic response are transferred into nonaffine relaxation channels, while one quarter remains in the coherent first harmonic:
\begin{equation}
    \frac{\chi_{\max}}
    {1-\chi_{\max}}
    =
    3.
\end{equation}

Hence,
\begin{equation}
    |\mathrm{NLI}|_{\max}=3
\end{equation}
emerges naturally from the interplay between
the eight-chain representation of the load-bearing network and the central-force nonaffine isostatic threshold in three dimensions, valid for fully-flexible chains, which provides the upper bound for the nonaffine correction $G_{NA}$.
This argument should be interpreted as a network-level mechanical compatibility relation rather than as a microscopic monomer-contact counting derivation.

\section{Alternative tube-orientation estimate of the limiting NLI}

The same limiting value can also be motivated directly from the tube-orientation stress of reptation theory \cite{deGennes1971,RubinsteinColby2003}. In the Doi--Edwards picture \cite{DoiEdwards1986}, the elastic shear stress carried by an entangled chain may be written schematically as
\begin{equation}
    \sigma_{xy}^{E}
    =
    G_N^0 \lambda^2 S_{xy},
\end{equation}
where \(G_N^0\) is the plateau modulus, \(\lambda\) is the chain stretch ratio, and
\begin{equation}
    S_{xy}
    =
    \langle u_xu_y\rangle
\end{equation}
is the shear component of the tube-orientation tensor.

At large deformation, chain stretch cannot grow indefinitely. Taking
\begin{equation}
    \lambda \rightarrow \lambda_{\mathrm{sat}},
\end{equation}
the maximum elastic stress is controlled by the maximum physically admissible value of \(S_{xy}\).

Since \(\bm{u}\) is a unit vector,
\begin{equation}
    u_x^2+u_y^2+u_z^2=1.
\end{equation}
The maximum value of the shear component is obtained when the tube segments lie in the shear plane and are oriented at \(45^\circ\), namely
\begin{equation}
    u_x=u_y=\frac{1}{\sqrt{2}},
    \qquad
    u_z=0.
\end{equation}
Hence
\begin{equation}
    S_{xy}^{\max}
    =
    u_xu_y
    =
    \frac{1}{2}.
\end{equation}

The corresponding orientation tensor is
\begin{equation}
    \bm{S}_{\max}
    =
    \begin{pmatrix}
    1/2 & 1/2 & 0\\
    1/2 & 1/2 & 0\\
    0 & 0 & 0
    \end{pmatrix}.
\end{equation}
Relative to the isotropic state,
\begin{equation}
    \bm{S}_{0}
    =
    \frac{1}{3}\bm{I},
\end{equation}
the deviatoric orientation tensor is
\begin{equation}
    \bm{Q}_{\max}
    =
    \bm{S}_{\max}
    -
    \frac{1}{3}\bm{I}
    =
    \begin{pmatrix}
    1/6 & 1/2 & 0\\
    1/2 & 1/6 & 0\\
    0 & 0 & -1/3
    \end{pmatrix}.
\end{equation}

Thus the maximum shear-orientation component is
\begin{equation}
    Q_{xy}^{\max}
    =
    \frac{1}{2},
\end{equation}
whereas the residual diagonal orientational component in the shear plane is
\begin{equation}
    Q_{xx}^{\max}
    =
    Q_{yy}^{\max}
    =
    \frac{1}{6}.
\end{equation}

The ratio between the saturated nonlinear shear-orientation contribution and the residual coherent orientational storage is therefore
\begin{equation}
    \frac{Q_{xy}^{\max}}{Q_{xx}^{\max}}
    =
    \frac{1/2}{1/6}
    =
    3.
\end{equation}

In the LAOS decomposition, the first harmonic represents the residual coherent phase-locked elastic storage, whereas the nonlinear elastic correction represents the higher-harmonic distortion generated by saturation of tube orientation. Therefore, in the fully saturated tube-orientation limit,
\begin{equation}
    \left|
    \frac{G'_{NL}}{G'_1}
    \right|_{\max}
    \simeq
    \frac{Q_{xy}^{\max}}{Q_{xx}^{\max}}
    =
    3.
\end{equation}

Since
\begin{equation}
    \mathrm{NLI}
    =
    -\frac{G'_{NL}}{G'_1}
\end{equation}
for strain-softening nonlinear elasticity, this gives
\begin{equation}
    \mathrm{NLI}_{\max}\simeq 3.
\end{equation}

This provides a complementary ``reptative'' interpretation of the limiting value. Once the tube orientation has reached its maximum admissible shear alignment, \(S_{xy}^{\max}=1/2\), further imposed deformation cannot increase the coherent phase-locked orientation. Instead, the additional response is transferred into nonlinear waveform distortion. The value \(\mathrm{NLI}_{\max}\simeq3\) therefore reflects the geometry of the saturated tube-orientation tensor.

This argument should be regarded as a molecular saturation estimate rather than an exact Fourier theorem. This molecular saturation picture is consistent with the constraint-counting result above. In the fully flexible-chain limit, bond-bending constraints no longer provide an additional source of rigidity, so the maximum nonaffine relaxation is controlled by the central-force isostatic threshold. At the same time, reptation theory shows that the tube-orientation tensor has a purely geometric upper bound: once the chains are maximally aligned in the shear plane, the shear component of orientation cannot grow further. Both arguments therefore describe the same physical limit: the coherent affine-like tube response has exhausted its ability to store additional elastic stress, and further deformation is converted into nonlinear, nonaffine waveform distortion. In this saturated state, only one quarter of the affine response remains in the first harmonic, while three quarters are transferred to nonlinear relaxation channels, giving the limiting estimate \(\mathrm{NLI}_{\max}\simeq3\).

\section{Conclusions}

We have developed a phenomenological framework connecting the nonlinear harmonic response observed in LAOS to the frequency-dependent nonaffine relaxation theory of disordered elasticity. The central result is a microscopic molecular interpretation of the nonlinear index (NLI) in terms of a competition between coherent affine-like tube elasticity and strain-dependent nonaffine relaxation associated with convective constraint release, tube destruction, and loss of orientational memory.

Within this framework, the first harmonic measures the residual phase-locked elastic storage of the entangled tube network, whereas higher harmonics quantify the progressive transfer of elastic response into nonlinear relaxation channels. The polymer analogue of the decreasing load-bearing connectivity in amorphous solids is identified as the survival fraction of tube constraints \(q\), while the tube orientation tensor \(S_{xy}\) describes the orientational state of those surviving constraints.

A key outcome of the theory is that the nonlinear response is governed not by universal fixed exponents, but by a crossover scaling form,
\begin{equation}
\mathrm{NLI}
=
\mathcal{F}
\left(
\frac{\gamma_0}{\gamma_c}
\right),
\qquad
\gamma_c
\sim
\frac{1}{\beta \Omega \tau_d},
\end{equation}
where \(\gamma_0\) is the strain amplitude, \(\tau_d\) is the disengagement time, and \(\beta\Omega\) characterizes the effective rate of nonlinear tube relaxation. The onset of nonlinear behavior is therefore controlled by the competition between externally imposed deformation and intrinsic tube-memory relaxation.

The proposed crossover model successfully reproduces the experimental NLI master curve across chemically and topologically different polymers using a minimal three-parameter form, yielding \(N_{\max}\simeq1.72\), a normalized crossover value \(\gamma_c\simeq39.2\) in the variable \(\gamma_0/\gamma_0^{LR}\), and an effective crossover exponent \(m\simeq1.58\). The fitted value \(N_{\max}\approx1.72\) is not imposed by the theory and remains below the ideal limiting value \(|\mathrm{NLI}|_{\max}=3\), indicating that the experiments probe a strongly nonlinear but not fully saturated constraint-collapse regime.
The fitted exponent \(m\simeq1.58>1\) further suggests a nonlinear constraint-collapse regime in which higher-harmonic generation and tube-constraint loss are amplified beyond the simplest CCR-controlled saturation scenario, possibly due to progressive tube dilation and heterogeneous constraint release during the oscillation cycle. In this sense, the present framework is consistent with molecular tube-theory approaches to nonlinear entangled polymer rheology and convective constraint release, originally introduced by Marrucci and later developed in quantitative tube models by Read and co-workers~\cite{Read2012}, as well as with Fourier-transform LAOS studies of linear and branched polymer architectures by Vlassopoulos and collaborators~\cite{HassagerVlassopoulos2008}.

A further implication of the framework is a possible constraint-counting interpretation of the limiting nonlinear index. In a Arruda--Boyce eight-chain representation, the saturated polymer network has eight effective load-bearing strands, while a fully flexible central-force network has a nonaffine threshold of six. Thus, at maximum nonlinear relaxation, roughly three quarters of the affine elastic response are transferred into nonaffine channels, leaving one quarter in the coherent first harmonic. This gives the estimate \(|\mathrm{NLI}|_{\max}\simeq3\). This should be viewed as a coarse-grained network-level argument, not as a microscopic monomer-contact counting theorem.

The finite-strain modulus entering this construction should be understood as a local or tangent modulus along the deformation trajectory, rather than as a single equilibrium linear-response modulus. This interpretation is consistent with elastoplastic and incremental nonaffine descriptions, in which the stress is updated through a strain-dependent local stiffness of the evolving microstructure. The LAOS extension of the affine--nonaffine decomposition therefore provides a natural framework for interpreting nonlinear harmonic generation as the spectral consequence of a progressively evolving load-bearing network.

More broadly, the present work establishes a direct bridge between tube-based polymer dynamics, LAOS harmonic analysis, and microscopic nonaffine lattice dynamics. In this perspective, the NLI emerges as a Fourier-resolved dynamic nonaffinity parameter that measures the progressive breakdown of load-bearing connectivity under oscillatory deformation. The framework therefore suggests that nonlinear harmonic generation may represent a universal spectral signature of dynamically evolving connectivity across a broad class of disordered soft materials, including entangled polymers, colloidal glasses, metallic glasses, and amorphous solids. 

The emergence of a fitted saturation level approaching \(N_{\max}\sim2\), together with the theoretically derived upper bound \(|\mathrm{NLI}|_{\max}=3\), suggests that nonlinear harmonic spectroscopy may provide direct quantitative access to the progressive exhaustion of load-bearing tube connectivity in entangled polymers.

\section{Acknowledgments}
A.Z. acknowledges funding from the European Research Council (Grant No.~101043968),
and the US Army Research Office (W911NF-22-2-0256). Useful discussions with Shi-Qing Wang, Valeriy Ginzburg and Oleg Gendelman are gratefully acknowledged.

\bibliographystyle{unsrt}
\bibliography{references}

\end{document}